\documentclass[letterpaper]{revtex4}
\usepackage[T1]{fontenc}
\usepackage[latin1]{inputenc}

\makeatletter

\providecommand{\LyX}{L\kern-.1667em\lower.25em\hbox{Y}\kern-.125emX\@}
\newcommand{\noun}[1]{\textsc{#1}}

\makeatother

\begin{document}

\title{Strict Holism in a Quantum Superposition of Macroscopic States}

\author{\textbf{J. Acacio de Barros} }

\thanks{E-mail: barros@csli.stanford.edu. On leave from Dept. de Fisica--ICE, UFJF,
Juiz de Fora, MG, Brazil. }

\author{\textbf{Patrick Suppes}}

\thanks{E-mail: suppes@ockham.stanford.edu}

\address{CSLI - Ventura Hall, Stanford University, Stanford, CA 94305-4115}

\date{\today}

\begin{abstract}
We show that some \( N \)-particle quantum systems are holistic, such that
the system is deterministic, whereas its parts are random. The total correlation
is not sufficient to determine the probability distribution, showing a need
for extra measurements. We propose a formal definition of holism not based on
separability. 
\end{abstract}
\maketitle

\section{Introduction}

In his famous elementary textbook, Richard Feynman claims that the only mystery
of quantum mechanics is exemplified by the electron self-interference in the
two-slit experiment \cite{Feynmann}. Interference is a consequence of the superposition
principle, and indeed most of the puzzling aspects of quantum mechanics are
related to the superposition of two or more states, as is the case in the Einstein-Podolsky-Rosen
(EPR) paradox \cite{EPR} or the Greenberger-Horne-Zeilinger (GHZ) theorem \cite{GHZ}.
Both EPR and GHZ show a striking characteristic of quantum mechanics: the nonseparability
of systems situated far apart from each other. In quantum mechanics, systems
that interacted with each other in the past may become entangled, and, even
if they are separated by a great distance later on, their properties can be
correlated in a way that would evade any attempt to give a classical explanation
\cite{Bell}. This nonseparability has as a consequence the nonexistence of
a joint probability distribution, and hence of a local hidden-variable theory,
that explains the outcome of the experiments \cite{suppeszannoti}. More recently,
Mermin \cite{Mermin} showed that if we allow states with a large number \( N \)
of particles to be superposed in a way similar to the superposition of particles
in the GHZ theorem, then quantum mechanics deviates exponentially with \( N \)
from the classical case (i.e., one that could be understood by a local hidden-variable). 

The nonexistence of local-hidden variables that can account for all the experimental
outcomes suggests that quantum mechanics has some holistic characteristic. Holism
is the idea that the whole cannot be considered as the sum of its individual
parts. The fact that systems far apart are nonseparable has led some authors
to suggest that quantum mechanics has in its core a holistic characteristic
\cite{Ghirardi,Primas}. Nonseparability, in the sense used in EPR or GHZ, means
that a local hidden-variable theory that predicts the outcome of the experiments
is impossible. Of course, nonseparability implies holism, but that the converse
is not true is what we show in this paper. To do this, we will first show that
a GHZ \( N \)-particle quantum mechanical system behaves in a deterministic
way, when considered as a whole, but that every proper subsystem of this system
behaves in a completely random way. This is done by first showing that any subsystem
has maximal entropy, whereas the whole system has entropy zero. Then, we analyze,
from a probabilistic point of view, the \( N \)-particle GHZ example. We show
that quantum mechanics is more restrictive on the subsystems than pure probability
considerations, even though, for the particular observables in question, a joint
probability distribution exists. Then, we propose a definition of holism that
is distinct from the concept of separability, and discuss this definition by
means of simple examples. Our definition of holism is satisfied by the GHZ quantum
mechanical system presented earlier.

\section{Quantum Mechanical Holism}

Let us start with the entangled GHZ-like \( N \)-particle state
\begin{equation}
\label{N-particle_GHZ}
|\psi \rangle =\frac{1}{\sqrt{2}}\left[ \prod ^{N}_{k=1}|+\rangle _{k}+\prod ^{N}_{k=1}|-\rangle _{k}\right] ,
\end{equation}
where \( \widehat{\sigma }_{iz}|+\rangle _{i}=|+\rangle _{i} \), \( \widehat{\sigma }_{iz}|-\rangle _{i}=-|-\rangle _{i} \),
with \( \widehat{\sigma }_{iz} \) being the spin operator in the \( z \) direction
acting on the \( i \)-th particle. It is easy to show that this state is an
eigenstate, with eigenvalue \( 1 \), of the observable operator
\begin{equation}
\hat{\Sigma }=\widehat{\sigma }_{1x}\otimes \widehat{\sigma }_{2x}\otimes \cdots \otimes \widehat{\sigma }_{Nx}.
\end{equation}
 In other words, the observable \( \hat{\Sigma } \), made out of the product
of all \( N \) spin observables, is deterministic, as a measurement of it always
results in the value \( 1. \) In a similar way, this determinism is also true
for the observables 
\begin{equation}
\prod _{i}\widehat{\sigma }_{iy}\otimes \prod _{j}\widehat{\sigma }_{jx},
\end{equation}
where the index \( i \) is any subset with even cardinality of \( 2^{\{1,2,\ldots ,N\}} \),
and \( j \) is the complement of \( i. \) 

The state (\ref{N-particle_GHZ}) has been the focus of several interesting
papers, all of them related to the deterministic aspects of the above observables
\cite{GHZ,Mermin,GHZ1,GHZ2,GHZ3,GHZ4,GHZ5,GHZ6}. However, in this paper we
will be interested in observables acting only on a subset of the set of all
particles in (\ref{N-particle_GHZ}). We start with the following. 

\begin{description}
\item [Proposition~1]Given the ket 
\begin{equation}
|\psi \rangle =\frac{1}{\sqrt{2}}(|++\cdots +\rangle +|--\cdots -\rangle ),
\end{equation}
and the spin operators \( \widehat{\sigma }_{id} \), where \( i=1\ldots N \)
and \( d=x,y,z, \) then any product of \( n<N \) distinct spin operators has
expectation zero. 
\end{description}
\emph{Proof.} Let us start with a hermitian operator \( \hat{\Sigma }^{'} \)
that is the product of \( n<N \) distinct spin operators, such that we can
write \( \hat{\Sigma }^{'} \) as 
\begin{equation}
\hat{\Sigma }^{'}=\prod ^{a}_{k=1}\hat{\sigma }_{k,x}\otimes \prod ^{b}_{k=a+1}\hat{\sigma }_{k,y}\otimes \prod ^{c}_{k=b+1}\hat{\sigma }_{k,z}\otimes \prod ^{N}_{k=c+1}\hat{1}_{k},
\end{equation}
 with \( 0<a<b<c<n \), and \( a+b+c=n \). We want to compute \( \langle \psi |\hat{\Sigma }^{'}|\psi \rangle , \)
the expected value of this operator, so \emph{
\begin{eqnarray}
\langle \psi |\hat{\Sigma }^{'}|\psi \rangle  & = & \frac{1}{2}i^{b-a-1}\left[ \prod ^{N}_{k=1}\langle +|_{k}+\prod ^{N}_{k=1}\langle -|_{k}\right] \times \nonumber \\
 &  & \left[ \prod ^{b}_{k=1}|-\rangle _{k}\prod ^{N}_{k=b+1}|+\rangle _{k}\right. \nonumber \\
 &  & \left. -\left( -1\right) ^{c+a}i^{a}\prod ^{b}_{k=1}|+\rangle _{k}\prod ^{N}_{k=b+1}|-\rangle _{k}\right] .
\end{eqnarray}
}From the equation above, it is immediate that the inner product is zero if
\( b<N, \) as we wanted to prove. 

Proposition 1 shows that the correlations for the \( N \)-particle system are
quite strange. We have a set of \( N \) particles that has always the same
observable associated to its totality, but when we look at any of its parts,
then the parts are completely uncorrelated. In this system the presence of a
nonzero correlation appears only when we look at the system as a whole, and
not at its parts. In the next section we will analyze in details the probabilistic
properties of the probability distribution associated to, say, the operator
\( \hat{\Sigma } \).

\section{Probabilistic Properties}

It is interesting to note the consequences of the previous result. Say we are
measuring the spin in the \( x \) direction for \( n<N \) particles. In this
case all the particles are independent, and also behave in a completely random
way, as the probability of measuring \( 1 \) is the same as the probability
of measuring \( -1 \). However, if we measure the spin of \emph{all} \( N \)
particles, the whole system is deterministic in a sense that will be made clear
later. First, let us start with the following Proposition. 

\begin{description}
\item [Proposition~2]Let
\begin{equation}
|\psi \rangle =\frac{1}{\sqrt{2}}(|++\cdots +\rangle +|--\cdots -\rangle ),
\end{equation}
 \( \hat{\Sigma }=\prod ^{N}_{k=1}\hat{\sigma }_{k,x}, \) and to each particle
\( i \), \( 1\leq i\leq N \), we associate the random variable \( \mathbf{S}_{i} \),
representing the value of its spin measurements, taking values \( \pm 1 \).
If \( t=n\Delta t \), \( n=0,1,2,\ldots  \) and we measure \( |\psi \rangle  \)
using \( \hat{\Sigma } \) at each \( t \). We define the random variables
\( \mathbf{X}^{\{k\}}_{t}=\prod _{\{k\}}\mathbf{S}_{k} \), where \( \{k\} \)
is any proper subset of \( \{1,\ldots ,N\} \) and \( \mathbf{X}_{t}=\prod ^{N}_{k=1}\mathbf{S}_{k} \).
Then each \( \mathbf{X}^{\{k\}}_{t} \), and \( \mathbf{X}_{t} \) define Bernouilli
processes. 
\end{description}
\emph{Proof.} First we should note that \( |\psi \rangle  \) is an eigenstate
of \( \hat{\Sigma } \), such that we can measure \( \hat{\Sigma } \) as many
times as we want without affecting \( |\psi \rangle  \). If we keep measuring
spin in the \( x \) direction for all particles in equal intervals of time
\( \Delta t \), we can make a data table for the experimental result that would
look like Table \ref{table1}, where we associate to each of the spin measurements
for particle \( i \) the random variable \( \mathbf{S}_{i} \) taking values
\( \pm 1 \). 
\begin{table}
{\centering \begin{tabular}{|c|c|c|c|c|c|}
\hline 
&
\( \mathbf{S}_{1} \)&
\( \mathbf{S}_{2} \)&
\emph{\( \cdots  \)}&
\( \mathbf{S}_{N} \)&
\( \prod _{i=1}^{N}\mathbf{S}_{i} \)\\
\hline 
\hline 
\( 0 \)&
\( 1 \)&
\( -1 \)&
\emph{\( \cdots  \)}&
\( 1 \)&
\( 1 \)\\
\hline 
\( \Delta t \)&
\( 1 \)&
\( -1 \)&
\emph{\( \cdots  \)}&
\( -1 \)&
\( 1 \)\\
\hline 
\( 2\Delta t \)&
\( -1 \)&
\( 1 \)&
\emph{\( \cdots  \)}&
\( 1 \)&
\( 1 \)\\
\hline 
\( 3\Delta t \)&
\( -1 \)&
\( -1 \)&
\emph{\( \cdots  \)}&
\( -1 \)&
\( 1 \)\\
\hline 
\( \vdots  \)&
\( \vdots  \)&
\( \vdots  \)&
&
\( \vdots  \)&
\( \vdots  \)\\
\hline 
\end{tabular}\par}

\caption{\label{table1}Possible set of experimental data results for the random variables
\protect\( \mathbf{S}_{1}\protect \), \protect\( \mathbf{S}_{2}\protect \),\emph{\protect\( \cdots \protect \)},
\protect\( \mathbf{S}_{N}\protect \), and \protect\( \prod _{i=1}^{N}\mathbf{S}_{i}\protect \).}
\end{table}
 Each column of this table would be completely uncorrelated to the any other
column or combinations of columns with less than \( N \) columns involved.
Similar independence and randomness hold for any row of length at most \( N-1 \),
i.e., at least one entry is deleted. However, if we multiply \( \mathbf{S}_{1} \),
\( \mathbf{S}_{2} \),\emph{\( \cdots  \)}, \( \mathbf{S}_{N} \), we always
obtain the same value \( \prod _{i=1}^{N}\mathbf{S}_{i}=1 \). Furthermore,
since the wave function \( |\psi \rangle  \) is unchanged, the equal probabilities
of obtaining a \( 1 \) or \( -1 \) for each of the columns or shortened rows
are also unchanged. As a consequence, the temporal sequence of product random
variables \( \mathbf{X}_{t}^{\{k\}}=\prod _{\{k\}}\mathbf{S}_{k} \), where
\( \{k\} \) is any proper subset of \( \{1,\ldots ,N\} \), form a Bernouilli
process, i.e. at each time \( t \) the random variables \( \mathbf{X}_{t}^{\{k\}} \)
are independently and identically distributed, as we wanted to show. It is straightfoward
to extend the same argument to \( \mathbf{X}_{t}. \)

We are now in a position to make explicit the statement that the system as a
whole is deterministic and its subsystems are random. 

\begin{description}
\item [Proposition~3]The random variables \( \mathbf{X}^{\{k\}}_{t}=\prod _{\{k\}}\mathbf{S}_{k} \),
where \( \{k\} \) is any proper subset of \( \{1,\ldots ,N\} \), defined in
a way similar to Proposition 2, have maximal entropy for such process, whereas
the random variable \( \mathbf{X}_{t}=\prod ^{N}_{k=1}\mathbf{S}_{k} \) has
zero entropy.  
\end{description}
\emph{Proof.} Since both \( \mathbf{X}^{\{k\}}_{t} \) and \( \mathbf{X}_{t} \)
define a Bernouilli process, their entropy is \( H=-\sum p_{i}\log p_{i} \),
where \( p_{i} \) is the probability of each possible outcome, in this case
\( \pm 1 \). \( \mathbf{X}_{t}=\prod _{i=1}^{N}\mathbf{S}_{i} \), representing
the system as a whole, has entropy zero, since for all \( t \) \( P(\mathbf{X}_{t}=1)=1 \)
and \( P(\mathbf{X}_{t}=-1)=0 \). Yet, any proper subset \( \{k\} \) of \( \{1,\ldots ,N\} \)
will define a random variable \( \mathbf{X}^{\{k\}}_{t}=\prod _{\{k\}}\mathbf{S}_{k} \)
whose entropy is maximal for such a process, as \( P(\mathbf{X}^{\{k\}}=1)=1/2 \)
and \( P(\mathbf{X}^{\{k\}}=-1)=1/2 \), i.e. the entropy \( H=-\sum p_{i}\log p_{i}=1 \),
where \( \log  \) is to base 2, as we wanted to prove. 

The results just obtained show that the system in question is strongly holistic,
in the sense that a measurement of \( \hat{\Sigma } \) containing \emph{all}
particles in the system yields a deterministic result, whereas any spin measurement
made on a subsystem has a perfectly random outcome. However, since we can measure
all the \( N \) spin values simultaneously, we can also write a data table
for the experimental outcomes, and a joint probability distribution exists.
In this sense, the system is holistic but is separable, as we can factor the
joint probability distribution.

Even though a joint probability distribution exists, we stress that such a strange
distribution, where only when we consider all particles is the system deterministic,
is rarely if ever found in any empirical domain. In fact, quantum mechanics
provides, as far as we know, the only example in nature of a case where we have
perfect correlation for a triple and zero correlation for pairs. This is the
case if we take a three-particle GHZ system, as it yields \( \mathbf{X}_{i} \)
\( \pm 1 \) random variables, with \( E(\mathbf{X}_{1}\mathbf{X}_{2}\mathbf{X}_{3})=1 \),
\( E(\mathbf{X}_{i})=0 \), \( i=1,\ldots ,3 \). It is also interesting to
stress that, in the three-particle GHZ case, the pair correlations are zero
as a consequence of the triple correlation and the individual expectations.
This can be verified by direct computation. \emph{}Say we have \( E(\mathbf{X}_{1}\mathbf{X}_{2}\mathbf{X}_{3})=1 \).
Then, all terms with 0 or 2 negative components sum to 1, i.e.,
\begin{equation}
x_{1}x_{2}x_{3}+\bar{x}_{1}\bar{x}_{2}x_{3}+\bar{x}_{1}x_{2}\bar{x}_{3}+x_{1}\bar{x}_{2}\bar{x}_{3}=1,
\end{equation}
where we use the notation \( x_{1} \)to represent \( P(\mathbf{X}_{1}=1) \),
\( \bar{x}_{1} \)to represent \( P(\mathbf{X}_{1}=-1) \), \( \bar{x}_{1}x_{2} \)
to represent \( P(\mathbf{X}_{1}=-1,\mathbf{X}_{2}=1) \), and so on. We also
have that
\begin{eqnarray}
x_{1}x_{2}=x_{1}x_{2}x_{3}=x_{1}x_{3}=x_{2}x_{3} & = & a,\label{system1} \\
\bar{x}_{1}\bar{x}_{2}=\bar{x}_{1}\bar{x}_{2}x_{3}=\bar{x}_{1}x_{3}=\bar{x}_{2}x_{3} & = & b,\\
\bar{x}_{1}x_{2}=\bar{x}_{1}x_{2}\bar{x}_{3}=\bar{x}_{1}\bar{x}_{3}=x_{2}\bar{x}_{3} & = & c,\\
x_{1}\bar{x}_{2}=x_{1}\bar{x}_{2}\bar{x}_{3}=x_{1}\bar{x}_{3}=\bar{x}_{2}\bar{x}_{3} & = & d,\label{system2} 
\end{eqnarray}
with \( a+b+c+d=1. \) Next, from (\ref{system1})--(\ref{system2}), \( x_{1}=a+d, \)
\( \bar{x}_{1}=b+c, \) \( x_{2}=a+c, \) \( \bar{x}_{2}=b+d, \) \( x_{3}=a+b, \)
\( \bar{x}_{3}=c+d \), and from \( E(\mathbf{X}_{i})=0 \), \( x_{1}=x_{2}=\bar{x}_{1}=\bar{x}_{2}=\frac{1}{2} \).
From (\ref{system1})--(\ref{system2}) and the following equations, we obtain
at once \( a=b=c=d \) and 
\begin{equation}
E(\mathbf{X}_{1}\mathbf{X}_{2})=E(\mathbf{X}_{2}\mathbf{X}_{3})=E(\mathbf{X}_{1}\mathbf{X}_{3})=0.
\end{equation}

However, contrary to the three-particle case, if we increase the number of particles
to four, the correlations are not dictated by \( E(\mathbf{X}_{1}\mathbf{X}_{2}\mathbf{X}_{3}\mathbf{X}_{4})=1 \),
\( E(\mathbf{X}_{i})=0 \), \( i=1,\ldots ,4 \) anymore. For the four-particle
case, we can compute, in a manner similar to the three-particle one, that \( E(\mathbf{X}_{i}\mathbf{X}_{j}\mathbf{X}_{k})=0 \),
\( i<j<k. \) However, the correlations \( E(\mathbf{X}_{i}\mathbf{X}_{j}) \)
can individually, but not independently, take any value in the closed interval
\( [-1,1] \). On the other hand, if all the correlations are zero, then the
positive atoms have a uniform distribution, by an argument similar to the one
given above. In fact, we can show the following.

\begin{description}
\item [Proposition~4]Given \( E(\mathbf{X}_{1}\cdots \mathbf{X}_{n})=0 \) and the
product of any nonempty subset of the random variables \( \mathbf{X}_{1}\cdots \mathbf{X}_{n} \)
also has expectation zero, including \( E(\mathbf{X}_{i})=0 \), \( 1\leq i\leq n \).
Then the \( 2^{n} \) atoms of the probability space supporting \( \mathbf{X}_{1}\cdots \mathbf{X}_{n} \)
has a uniform probability distribution, i.e., each atom has probability \( 1/2^{n}. \) 
\end{description}
\emph{Proof.} We show this by induction. For \( n=1 \), we have by hypothesis
that \( E(\mathbf{X}_{i})=0 \), so, as required, \( P(\mathbf{X}_{i}=1)=x_{1}=1/2. \)
Next, our inductive hypothesis is that for every subsystem having \( m<n \),
the \( 2^{m} \) atoms have a uniform distribution, and we need to show this
holds for \( n \). Using the induction hypothesis for \( n-1 \), we have at
once the following pair of equations: 
\begin{eqnarray*}
x_{1}x_{2}\cdots x_{n-1} & = & x_{1}x_{2}\cdots x_{n-1}x_{n}+x_{1}x_{2}\cdots x_{n-1}\bar{x}_{n}=2^{1-n},\\
x_{1}x_{2}\cdots x_{n-2}x_{n} & = & x_{1}x_{2}\cdots x_{n-1}x_{n}+x_{1}x_{2}\cdots \bar{x}_{n-1}x_{n}=2^{1-n}.
\end{eqnarray*}
Subtracting one equation from the other we have at once \( x_{1}x_{2}\cdots \bar{x}_{n-1}x_{n}=x_{1}x_{2}\cdots x_{n-1}\bar{x}_{n}. \)
By similar arguments, we show that all atoms that have exactly one negative
value of \( \bar{x}_{i} \) for the \( n \)-particle case are equal in probability.
Moreover, without any new complication this argument extends to equal probability
for any atom having exactly \( k \) negative values, \( 2\leq k\leq n \). 

Next, we can easily show that those atoms differing by 2, and therefore by an
even number of, negative values have equal probability. We give the argument
for \( k=0 \) and \( k=2 \):
\begin{eqnarray*}
x_{1}x_{2}\cdots x_{n-1} & = & x_{1}x_{2}\cdots x_{n-1}x_{n}+x_{1}x_{2}\cdots x_{n-1}\bar{x}_{n}=2^{1-n},\\
\bar{x}_{1}x_{2}\cdots x_{n-2}x_{n} & = & \bar{x}_{1}x_{2}\cdots x_{n-1}x_{n}+\bar{x}_{1}x_{2}\cdots \bar{x}_{n-1}x_{n}=2^{1-n}.
\end{eqnarray*}
Using the previous result and subtracting we get \( x_{1}x_{2}\cdots x_{n-1}x_{n}=\bar{x}_{1}x_{2}\cdots x_{n-1}\bar{x}_{n}. \)
Finally, we use the hypothesis that \( E(\mathbf{X}_{1}\cdots \mathbf{X}_{n})=0 \).
This zero expectation requires that the sum of all the terms with 0 or an even
number of negative values have the same sum as all the terms with an odd number
of negative values. This implies at once that all atoms have equal probability,
and so each has probability \( 1/2^{n} \), proving Proposition 4. 

We also prove a more restricted result, but a sifnificant one, by purely probabilistic
means, i.e., no quantum mechanical concepts or assumptions are needed in the
proof. 

\begin{description}
\item [Proposition~5]Given \( E(\mathbf{X}_{1}\ldots \mathbf{X}_{N})=\pm 1 \) and
\( E(\mathbf{X}_{i})=0 \), \( i=1,\ldots ,N \), then any correlation of \( N-1 \)
particles is zero, e.g., \( E(\mathbf{X}_{1}\ldots \mathbf{X}_{N-1})=0 \),
\( E(\mathbf{X}_{1}\ldots \mathbf{X}_{N-2}\mathbf{X}_{N})=0 \), etc. 
\end{description}
\emph{Proof.} We give the proof for \( E(\mathbf{X}_{1}\ldots \mathbf{X}_{N})=1 \).
Then there are \( 2^{N} \) atoms in the probability space. Given the expectation
equal to 1, half ot the atoms must have probability 0, namely all those representing
negative spin products. Now, we consider all the terms expressing \( E(\mathbf{X}_{1}\ldots \mathbf{X}_{N-1}) \).
On the positive side, we have all those with even or zero negative values:
\begin{equation}
\label{expansion}
x_{1}x_{2}\cdots x_{N-1}+\bar{x}_{1}\bar{x}_{2}\cdots x_{N-1}+\cdots +\bar{x}_{1}\bar{x}_{2}\cdots \bar{x}_{N-1}
\end{equation}
if \( N-1 \) is even and as the last term if \( N-1 \) is odd \( x_{1}\bar{x}_{2}\cdots \bar{x}_{N-1} \).
To be extended to atoms, a positive \( x_{N} \) must be added. So, in probability
\[
x_{1}x_{2}\cdots x_{N-1}=x_{1}x_{2}\cdots x_{N-1}x_{N},\]
because, given \( E(\mathbf{X}_{1}\ldots \mathbf{X}_{N})=1 \)
\[
x_{1}x_{2}\cdots x_{N-1}\bar{x}_{N}=0,\]
and similar for the other terms in (\ref{expansion}).

The same thing applies in similar fashion to the negative side, e.g., 
\[
\bar{x}_{1}x_{2}\cdots x_{N-1}=\bar{x}_{1}x_{2}\cdots x_{N-1}\bar{x}_{N},\]
since the atom on the right must have zero or an even number of negative values. 

But we observe that, by hypothesis, \( E(\mathbf{X}_{1}\ldots \mathbf{X}_{N-2}\mathbf{X}_{N})=0 \),
but the probability \( x_{N} \) is just equal to the sum of the probabilities
of the positive terms of \( E(\mathbf{X}_{1}\ldots \mathbf{X}_{N-2}\mathbf{X}_{N}) \)
and \( \bar{x}_{N} \) is just equal to the sum of the probabilities of the
negative terms above. Since, \( x_{N}-\bar{x}_{N}=0 \), we conclude \( E(\mathbf{X}_{1}\ldots \mathbf{X}_{N-1})=0 \).
The same argument can be extended to the other \( N-1 \) combinations of \( \mathbf{X}_{i} \),
and this completes the proof.

\section{\protect\( \Pi \protect \)-Holism}

The remarkable property that a quantum system has a perfect correlation for
its whole but a totally random behavior for \emph{any} of its part seems to
us to represent a holistic characteristic of quantum mechanics. This holism
is, however, quite distinct from what is known in the literature as separability.
For that reason, we propose the following definition for strict holism. 

\begin{description}
\item [Definition]Let \( \Omega =(\Omega ,\cal {F},P) \) be a finite probability
space and let \( \mathbf{F}=\left\{ \mathbf{X}_{i},1\leq i\leq N\right\}  \)
be a family of \( \pm 1 \) random variables defined on \( \Omega  \). Let
\( \Pi  \) be a property defined for finite families of random variables. Then
\( \mathbf{F} \) is strictly \( \Pi  \)-holistic iff\\
\textbf{(i)} \( \mathbf{F} \) has \( \Pi  \);\\
\textbf{(ii)} No subfamily of \( \mathbf{F} \) has \( \Pi  \).\\
Moreover, if \( \Pi  \) is a numerical property, \\
\textbf{(iii)} No subfamily of \( \mathbf{F} \) approximates \( \Pi  \).
\end{description}
To understand this definition, let us give some examples from classical mechanics.
It is well know in classical gravitation theory that a two-particle system has
a well defined solution. However, if we add to this system an extra particle,
no closed solutions to this system exist in some cases, and in fact its behavior
can be completely random \cite{Alekseev}. One may be tempted to think that
this chaotic behavior is a holistic property, but according to the definition
above, it is not. For instance, let us take the restricted three-body problem
analyzed by Alekseev, where two particles with large mass orbit around their
Center of Mass (CM), while a third small particle oscillates in a line passing
through the CM and perpendicular to the plane of orbit of the two large masses.
The whole system behaves randomly, as well at least one subsystem, the one defined
by the small particle. Hence, this system is not \( \Pi  \)-holistic, if we
choose \( \Pi  \) to be the property of being random. 

As yet another example, let us consider a glass of water. The water is a large
system that does not behave like a water molecule, but in a coordinated way
dictated by hydrodynamics. Is then this system holistic? If we take, say, half
the glass of water, the properties of this half of water are the same as the
whole glass, except its mass, hence the system is not \( \Pi  \)-holistic for
the other macroscopic properties of the water. What about properties like, say,
mass? Say we take the full glass and remove only a water molecule from it. The
new subsystem approximates the mass of the original one, violating hypothesis
\textbf{(iii)} from the Definition, and so if we choose \( \Pi  \) to be the
property mass, the system is not \( \Pi  \)-holistic. 

\begin{description}
\item [Proposition~6]Let \( \mathbf{F}=\left\{ \mathbf{S}_{i},i=1,\ldots ,N\right\}  \)
be the set of random variables of all the spin measurements of the state 
\[
|\psi \rangle =\frac{1}{\sqrt{2}}(|++\cdots +\rangle +|--\cdots -\rangle ),\]
and let \( \mathbf{X}_{t} \) be the product random variable of Proposition
3, and let \( \mathbf{X}^{\{k\}}_{t} \) be the product random variable of any
subfamily \( \{k\} \) as defined earlier. Let the entropy be the \( \Pi  \)
property of these product random variables. Then \( \mathbf{F} \) is \( \Pi  \)-holistic. 
\end{description}
\emph{Proof.} Immediate, from Proposition 3, since the entropy of \( \mathbf{X}_{\mathbf{t}} \)
is \( 0 \) and, for any \emph{\( \{k\} \),} the entropy of \( \mathbf{X}^{\{k\}}_{\mathbf{t}} \)
is \( 1 \).

\section{Final Remarks}

To summarize, we found that an \( N \)-particle GHZ state has a strong holistic
property. However, it may be difficult to detect experimentally a quantum mechanical
holistic characteristic with a large number of particles, as decoherence may
play an important role, given that the decoherence time decreases rapidly if
we increase the number of particles \cite{decoherence1,decoherence2,decoherence3}.
A promising setup where this holism could be verified for a reasonably large
number of particles is the one proposed by Cirac and Zoller \cite{Garg,Garg2}.
We found that for \( N\geq 4 \), the measurements of \( E(\mathbf{X}_{1}\mathbf{X}_{2}\mathbf{X}_{3}\cdots \mathbf{X}_{N}) \)
and of \( E(\mathbf{X}_{i}) \) do not fix a probability distribution, and extra
measurements are necessary for the pairs, triples, and so on, for the probability
distribution to be fixed. We believe that these measurements, which should yield
many zero correlations, could be used to put additional constraints on some
local-hidden variable models that exploit the detection loophole \cite{detectionloophole,loophole2,loophole3,loophole4}.

\end{document}